\documentclass[12pt]{article}
\usepackage{amssymb,epsfig}

\newcommand{\pa}{\partial}
\newcommand{\al}{\alpha}

\renewcommand{\thefootnote}{\fnsymbol{footnote}}

\begin{document}

\title{
\begin{flushright}
\ \\*[-80pt] 
\begin{minipage}{0.25\linewidth}
\normalsize
hep-ph/0703044 \\
YITP-07-12 \\
KUNS-2065 \\*[50pt]
\end{minipage}
\end{flushright}
{\Large \bf 
Relaxed fine-tuning in models with non-universal gaugino masses
\\*[20pt]}}

\author{Hiroyuki~Abe$^{1,}$\footnote{
E-mail address: abe@yukawa.kyoto-u.ac.jp}, \ 
Tatsuo~Kobayashi$^{2,}$\footnote{
E-mail address: kobayash@gauge.scphys.kyoto-u.ac.jp} \ and \ 
Yuji~Omura$^{3,}$\footnote{
E-mail address: omura@scphys.kyoto-u.ac.jp}\\*[20pt]
$^1${\it \normalsize 
Yukawa Institute for Theoretical Physics, Kyoto University, 
% } \\ {\it \normalsize 
Kyoto 606-8502, Japan} \\
$^2${\it \normalsize 
Department of Physics, Kyoto University, 
Kyoto 606-8502, Japan} \\
$^3${\it \normalsize 
Department of Physics, Kyoto University, 
Kyoto 606-8501, Japan} \\*[50pt]}

\date{
\centerline{\small \bf Abstract}
\begin{minipage}{0.9\linewidth}
\medskip 
\medskip 
\small
We study, in a bottom-up approach, the fine-tuning problem 
between soft SUSY breaking parameters and the $\mu$-term 
for the successful electroweak symmetry breaking in the minimal 
supersymmetric standard model. It is shown that certain 
nontrivial ratios between gaugino masses, that is 
non-universal gaugino masses, are necessary at the GUT scale, 
in order for the fine-tuning to be reduced above $10$\,\% order. 
In addition, when all the gaugino masses should be regarded 
as independent ones in their origins, a small gluino mass 
$M_3 \lesssim 120$ GeV and a non-vanishing $A$-term $A_t \sim O(M_3)$ 
associated to top squarks are also required at the GUT scale 
as well as the non-universality. 
On the other hand, when we consider some UV theory, 
which fixes ratios of soft SUSY breaking parameters 
as certain values with the overall magnitude, 
heavier spectra are allowed.
It is favored that the gluino and wino masses are almost 
degenerate at the weak scale, while wider region of 
bino mass is favorable. 
\end{minipage}
}

\begin{titlepage}
\maketitle
\thispagestyle{empty}
\end{titlepage}

%\tableofcontents

\renewcommand{\thefootnote}{\arabic{footnote}}
\setcounter{footnote}{0}

\section{Introduction}
\label{sec:intro}
Supersymmetric extension of the standard model (SM) is one of the 
most promising candidates for a new physics at the TeV scale. 
It can stabilize the huge hierarchy between the electroweak 
(EW) scale and the Planck scale.
In particular, the minimal supersymmetric standard model (MSSM) 
is interesting from the viewpoint of its minimality.
Also the MSSM unifies three 
gauge couplings of SM gauge interactions at the grand unified 
theory (GUT) scale $M_{GUT} \sim 2 \times 10^{16}$ more precisely.
Furthermore, supersymmetric standard models 
provide sources for the dark matter.

Among such attractive features, the most remarkable one would be 
the radiative EW symmetry breaking~\cite{Inoue:1982ej}. 
The MSSM can automatically break EW symmetry due to the large 
logarithmic correction to the soft supersymmetry (SUSY) breaking 
mass $m_{H_u}$ for the up-sector Higgs field~\cite{Okada:1990gg}, 
\begin{eqnarray}
\Delta m_{H_u}^2 &\sim& 
-\frac{3y_t^2}{4\pi^2} m_{\tilde{t}}^2 
\ln \frac{\Lambda}{m_{\tilde{t}}}, 
\nonumber
\end{eqnarray}
which determines the size of $Z$-boson mass $M_Z$ as 
\begin{eqnarray}
\frac{1}{2}M_Z^2 &\sim& -\mu^2-m_{H_u}^2, 
\nonumber
\end{eqnarray}
through a minimization condition for the Higgs potential. 
Here, $y_t$ is the top Yukawa coupling, $m_{\tilde{t}}$ 
is the top squark mass, $\Lambda$ is the cut-off scale, 
and $\mu$ is the SUSY mass of up- and down-sector Higgs fields. 
We have assumed a (moderately) large value of 
$\tan \beta = \langle H_u \rangle / \langle H_d \rangle$.

On the other hand, the MSSM predicts the lightest CP-even 
Higgs mass at one-loop level, 
\begin{eqnarray}
m_h^2 &\le& 
M_z^2  + \frac{3 m_t^4}{4 \pi^2 v^2} 
\ln \frac{m_{\tilde{t}^2}}{m_t^2} + \cdots.
\nonumber
\end{eqnarray} 
The experimental bound $m_h \ge 114.4$ GeV requires 
$m_{\tilde{t}} \gtrsim 500$ GeV.
This value of $m_{\tilde{t}}$ leads to quite large 
correction $\Delta m_{H_u}^2$.
Thus, to obtain $M_Z$, we need typically a few percent fine-tuning between the 
SUSY mass $\mu$ and the soft SUSY breaking mass $m_{H_u}$ 
at the GUT scale, which are not related to each other in general. 
This is sometimes called a `little hierarchy problem'~\cite{Barbieri:1987fn}. 
There have been several works recently addressing this 
issue ~\cite{Brignole:2003cm}-\cite{Dermisek:2006qj}.
%~\cite{Batra:2003nj}Harnik:2003rs,Kobayashi:2004pu,Choi:2005uz,Choi:2005hd,Dermisek:2006ey,\cite{Dermisek:2006qj}). 
Most of them, however, are based on some specific models.

Here, we study the fine-tuning problem from the bottom-up viewpoint, 
and show what kind of model can relax this sort 
of fine-tuning. We will take two 
kinds of stances. One is a complete bottom-up approach, where 
all the soft SUSY breaking parameters are considered as 
independent ones to each other in their origins. In this case, 
we have to care about the sensitivity of the EW scale ($Z$-boson mass) 
to all the parameters at the GUT scale. The other is, 
in a sense, a half top-down approach. 
We suppose  some 
ultra-violet (UV) theories which fix certain ratios between 
the soft parameters at the GUT scale. 
Then we consider the 
fine-tuning between the remaining independent ones.
We will show preferable values of the ratios between 
gaugino masses and the $A$-term. 
Indeed, several models lead to non-universal gaugino masses 
as well as non-universal scalar masses and A-terms, e.g. 
moduli mediation \cite{Brignole:1993dj}, 
anomaly mediation \cite{Randall:1998uk}, 
mirage mediation \cite{Choi:2004sx,Choi:2005uz} and 
the SUSY breaking scenario, where F-components of gauge non-singlets 
are dominant \cite{Dermisek:2006qj,Ellis:1984bm}.
(See also Ref.~\cite{Choi:2007ka} for several classes of models leading to 
non-universal gaugino masses with certain ratios.)
%In each model, ratios among non-universal gaugino masses 
%are fixed as certain values.
Scalar masses and $A$-terms are more model-dependent.
However, in each model, ratios of gaugino masses and 
scalar masses as well as $A$-terms are fixed as certain values.
In these models, the independent parameter for SUSY breaking terms 
corresponds to the overall magnitude of SUSY breaking, say $M$, 
and we should concentrate to only the fine-tuning of the overall 
magnitude $M$.

The sections of this paper are organized as follows. In 
Section~\ref{sec:mssmft}, we briefly review the fine-tuning 
problem in the MSSM, and introduce fine-tuning parameters. 
In Section~\ref{sec:bottomup}, we discuss how the fine-tuning 
can be reduced when all the soft SUSY breaking parameters are 
regarded as independent ones. In Section~\ref{sec:uvmodel}, 
on the other hand, we examine the fine-tuning problem under the 
assumption that certain ratios between soft parameters, especially 
between gaugino masses, are fixed by some UV theories and find 
preferable ratios which reduce the fine-tuning. 
Section~\ref{sec:concl} is devoted to conclusions and discussions.
%In Appendix~\ref{app:mssmrge}, we show MSSM one-loop renormalization 
%group equations (RGE)~\cite{Inoue:1982ej} and integrate them 
%following Ref.~\cite{Ibanez:1983di}, when the Yukawa couplings and 
%the $A$-terms are neglected except for ones associated to the top 
%quark supermultiplets. 

\section{Fine-tuning problem in MSSM}
\label{sec:mssmft}
In this section we review the fine-tuning problem 
in the MSSM shortly, and then introduce fine-tuning 
parameters describing the sensitivity of the EW sale to 
the soft parameters at the GUT scale.  

The MSSM Higgs sector is described by the superpotential, 
\begin{eqnarray}
W_{SUSY} &=& \mu H_u H_d + y_t Q_3 U_3 H_u, 
\nonumber
\end{eqnarray}
and the relevant soft SUSY breaking terms are written as, 
\begin{eqnarray}
V_{\rm solf} &=& 
m_{Hu}^2 |H_u|^2 
+ m_{Hd}^2 |H_d|^2 +m_{Q_3}^2+m_{U_3}^2 
+ (\mu B H_u H_d + y_t A_t Q_3 U_3 H_u 
+ \textrm{h.c.}), 
\nonumber
\end{eqnarray}
where %$\mu$ is the SUSY mass ($\mu$-term) between 
%up-type and down-type Higgs fields $H_u$ and $H_d$, respectively, 
%$y_t$ is the top Yukawa coupling, $m_{H_u}$ ($m_{H_d}$) 
$m_{H_d}$, $m_{Q_3}$ and $m_{U_3}$ are the 
soft scalar mass for $H_d$, $Q_3$ and $U_3$, respectively, 
$\mu B$ is the SUSY breaking 
mass ($\mu B$-term) between $H_u$ and $H_d$, and $A_t$ is the 
scalar trilinear coupling ($A$-term) involving the top squarks. 
Throughout this paper, we neglect all the Yukawa couplings and the 
$A$-terms except for ones associated to the top quark supermultiplets, 
$y_t$ and $A_t$. Note that we use the same notation for denoting 
a chiral superfield and its lowest scalar component. 

The EW symmetry breaking causes the $Z$-boson mass 
$M_Z = 91.2$ GeV. 
A minimization condition of the total Higgs potential results 
in the following relation, 
\begin{eqnarray}
\frac{1}{2} M_Z^2 &=& 
-\mu^2(M_Z) 
-\frac{m_{Hu}^2(M_Z) \tan^2 \beta -m_{H_d}^2(M_Z)}{\tan^2 \beta-1} 
\nonumber \\ &\sim& 
-\mu^2(M_Z) -m_{Hu}^2(M_Z), 
\label{eq:zmass}
\end{eqnarray}
where and hereafter we assume a (moderately) large value of 
$\tan \beta=\langle H_u \rangle / \langle H_d \rangle$ like 
$\tan \beta \gtrsim 5$. 
The radiative correction to $m_{H_u}^2$ is dominantly given by 
the contributions from top squarks with mass scale $m_{\tilde{t}}$, 
which is estimated as 
\begin{eqnarray}
\Delta m_{H_u}^2(M_Z) &\approx& 
-\frac{3y_t^2(M_Z)}{4\pi^2} m_{\tilde{t}}^2 
\ln \frac{\Lambda}{m_{\tilde{t}}}. 
\label{eq:mhudominant}
\end{eqnarray}

On the other hand, within the two-loop approximation the lightest 
Higgs boson mass is constrained by~\cite{Carena:1995wu} 
\begin{eqnarray}
m_h^2 &\le& 
M_z^2 \cos^2 2\beta 
\Big( 1-\frac{3m_{t}^2}{8 \pi^2 v^2} 
\ln \frac{m_{\tilde{t}}^2}{m_t^2} \Big) 
\nonumber \\ &&
+ \frac{3 m_t^4}{4 \pi^2 v^2} 
\bigg[ 
\ln \frac{m_{\tilde{t}}^2}{m_t^2} 
+ \frac{\tilde{A_t}^2}{m_{\tilde{t}}^2}
\Big( 1-\frac{\tilde{A_t}^2}{12m_{\tilde{t}}^2}) 
\nonumber \\ &&
+ \frac{1}{16 \pi^2}
\Big( \frac{3m_t^2}{2v^2}-32 \pi \al_3 \Big) 
\bigg\{ \frac{2\tilde{A_t}^2}{m_{\tilde{t}}^2}
\Big( 1-\frac{\tilde{A_t}^2}{12m_{\tilde{t}}^2} \Big) 
\ln \frac{m_{\tilde{t}}^2}{m_t^2} 
+\Big( \ln \frac{m_{\tilde{t}}^2}{m_t^2} \Big)^2 \bigg\} 
\bigg], 
\label{eq:higgsmassbound}
\end{eqnarray}
where $\tilde{A_t}=A_t(M_Z)-\mu \cot \beta 
\approx A_t(M_Z)$ and $m_{\tilde{t}}$ is the 
averaged top squark mass, 
\begin{eqnarray}
m_{\tilde{t}}^2 
&=& \sqrt{m_{Q_3}^2(M_Z)\,m_{U_3}^2(M_Z)}. 
\label{eq:avrgstopmass}
\end{eqnarray}
The strong gauge coupling $g_3$, the vacuum value of the 
lightest Higgs field $v$, and the running top quark mass 
$m_t$ at the $M_Z$ scale are given by 
$\al_3(M_Z)=g_3^2/4\pi \approx 0.12$, 
$v=173.7$ GeV, and $m_t=164.5$ GeV, respectively. 

\begin{figure}[t]
\begin{center}
\begin{minipage}{0.5\linewidth}
\epsfig{file=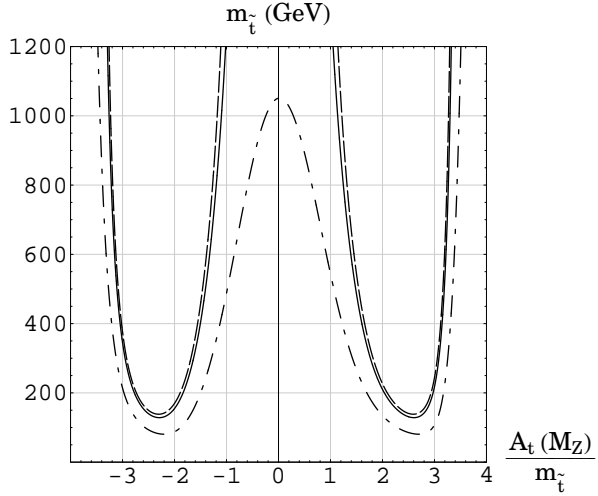,width=\linewidth} \\*[-40pt]
\end{minipage} 
\end{center}
\caption{The lower bound on the averaged top squark mass 
$m_{\tilde{t}}$ for $m_h \ge 114.4$ GeV (solid line) 
as well as $m_h \ge 110$ GeV (dot-dashed line) 
and $m_h \ge 115$ GeV (dashed line). 
The other parameters are chosen as 
$m_t=164.5$ GeV, $\mu=200$ GeV and $\tan \beta=10$.}
\label{fig:atmt}
\end{figure}

{}From the two-loop expression (\ref{eq:higgsmassbound}) 
and the observed lower bound by the LEP experiment 
$m_h^2 \ge 114.4$ GeV, we can estimate the allowed lowest value 
of the top squark mass which is shown in Fig.~\ref{fig:atmt}. 
{}From this figure it is apparent that a relatively large $A$-term 
at the EW scale 
\begin{eqnarray}
|A_t(M_Z)/m_{\tilde{t}}| &\gtrsim& O(1), 
\nonumber
\end{eqnarray} 
is favorable for the Higgs boson mass above the LEP bound. 
Furthermore, for a small value of $|A_t(M_Z)/m_{\tilde{t}}|$, 
a considerably large top squark mass is required as 
\begin{eqnarray}
m_{\tilde{t}} &\gtrsim& 500 \ \textrm{GeV}, 
\qquad 
{\rm for~~}|A_t(M_Z)/m_{\tilde{t}}| \lesssim 1.5, \nonumber \\ 
m_{\tilde{t}} &\gtrsim& 1000 \ \textrm{GeV}, 
\qquad 
{\rm for~~}|A_t(M_Z)/m_{\tilde{t}}| \lesssim 1.0.
\label{eq:mttlb}
\end{eqnarray}
This large top squark mass causes $m_{Hu}^2(M_Z)$ in 
Eq.~(\ref{eq:zmass}) to be much larger than $O(M_Z^2)$, 
because of the one-loop effect (\ref{eq:mhudominant}) 
with a large logarithm. 
Thus, $\mu^2$ must be fine-tuned in order to obtain the 
successful EW breaking with $M_Z \sim 91.2$ GeV. 
This is the so-called little hierarchy problem. 

The expressions (\ref{eq:zmass}), (\ref{eq:mhudominant}) 
and (\ref{eq:higgsmassbound}) are all written in terms of 
the low energy values of parameters such as $m_{H_u}^2(M_Z)$ 
and $m_{\tilde{t}}^2$. 
We express the soft 
parameters at the EW scale in terms of ones at the GUT 
scale~\cite{Ibanez:1983di}, by integrating 
the one-loop renormalization group equations~\cite{Inoue:1982ej}. 
For example, the gaugino masses at the EW scale are written 
in terms of themselves at the GUT scale as 
\begin{eqnarray}
M_1(M_z) &=& 0.41 M_1, 
\nonumber \\
M_2(M_z) &=& 0.82 M_2, 
\nonumber \\
M_3(M_z) &=& 2.91 M_3. 
\label{eq:maitogsp}
\end{eqnarray}
On the other hand, the scalar masses 
$m_{H_u}$, $m_{Q_3}$, $m_{U_3}$ 
and $A_t$ at the EW scale are given by  
\begin{eqnarray}
-2m_{Hu}^2(M_z) 
&=& 
5.45 M_3^2 +0.0677 M_3 M_1 -0.00975 M_1^2  
\nonumber \\ &&
+0.470 M_2 M_3 +0.0135 M_1 M_2  -0.433 M_2^2 
\nonumber \\ &&
+0.773 A_t M_3 +0.168 A_t M_2 +0.0271 A_t M_1 
\nonumber \\ &&
+0.214 A_t^2 -1.31 m_{Hu}^2 +0.690 m_{Q_3}^2 +0.690 m_{U_3}^2, 
\label{eq:mhuitogsp} \\
m_{Q_3}^2(M_Z) 
&=& 5.76 M_3^2 -0.0113 M_1 M_3 -0.00679 M_1^2  
\nonumber \\ &&
-0.0782 M_2 M_3 -0.00225 M_1 M_2 +0.400 M_2^2 
\nonumber \\ &&
-0.129 A_t M_3 +0.0281 A_t M_2 +0.00451 A_t M_1 
\nonumber \\ &&
-0.0357 A_t^2 -0.115 m_{H_u}^2 
+0.885 m_{Q_3}^2 -0.115 m_{U_3}^2, 
\label{eq:mqitogsp} \\
m_{U_3}^2(M_Z) 
&=& 4.85 M_3^2 -0.0226 M_1 M_3 +0.0453 M_1^2 
\nonumber \\ &&
-0.156 M_2 M_3 -0.00451 M_1 M_2 -0.183 M_2^2 
\nonumber \\ &&
-0.258 A_t M_3 +0.0561 A_t M_2 +0.00903 A_t M_1 
\nonumber \\ &&
-0.0713 A_t^2 -0.230 m_{H_u}^2 -0.230 m_{Q_3}^2 + 0.770 m_{U_3}^2, 
\label{eq:muitogsp} \\
A_t(M_Z) &=& 2.16 M_3 +0.268 M_2 +0.0340 M_1 +0.310 A_t. 
\label{eq:atitogsp}
\end{eqnarray}
Here the soft parameters without an argument in the right-hand 
side stand for the values at the GUT scale. We impose the 
boundary conditions $5\alpha_1/3=\alpha_2=\alpha_3=1/24$ at the 
GUT scale $M_{GUT}=2 \times 10^{-16}$ GeV and $y_t(M_Z)=m_t/v$ 
at $M_Z$. The $\mu$-parameter 
receives a small radiative correction, and is shown to be 
\begin{eqnarray}
\mu^2(M_Z) &=& 1.09 \mu^2. 
\label{eq:mutermitogsp}
\end{eqnarray}
%See Appendix~\ref{app:mssmrge} for some details of the derivation. 

The large contribution to the Higgs soft mass 
(\ref{eq:mhudominant}) from top squarks is now translated into 
the gluino mass squared $M_3^2$ with the largest coefficient 
$5.45$ in Eq.~(\ref{eq:mhuitogsp}). The mass squared $M_3^2$ 
also appears in $m_{Q_3}^2(M_Z)$ and $m_{U_3}^2(M_Z)$ in Eqs.~(\ref{eq:mqitogsp}) 
and (\ref{eq:muitogsp}), respectively, as dominant terms. From 
Eqs.~(\ref{eq:avrgstopmass}), (\ref{eq:mqitogsp}) and (\ref{eq:muitogsp}), 
if all the soft parameters take similar values, i.e., 
$M_a \approx m_i \approx A_t$ ($a=1,2,3$), we find 
\begin{eqnarray}
m_{\tilde{t}}^2 &\approx& 5 M_3^2. 
\label{eq:mttitom3}
\end{eqnarray}
{}From Eqs.~(\ref{eq:mttlb}) and (\ref{eq:mttitom3}), the lower 
bound for $M_3$ is estimated as 
\begin{eqnarray}
M_3 &\gtrsim& 220 \ \textrm{GeV}, \qquad 
{\rm for~~}|A_t(M_Z)/m_{\tilde{t}}| \lesssim 1.5, \nonumber \\
M_3 &\gtrsim& 450 \ \textrm{GeV}, \qquad 
{\rm for~~}|A_t(M_Z)/m_{\tilde{t}}| \lesssim 1.0, 
\label{eq:m3lb}
\end{eqnarray}
in order to satisfy the Higgs mass bound (\ref{eq:higgsmassbound}). 
Thus $M_3^2$ term with the large coefficient in 
Eq.~(\ref{eq:mhuitogsp}) and then in Eq.~(\ref{eq:zmass}) 
is much larger than $M_Z^2$. The other terms such as $\mu$ in the 
right-hand side of Eq.~(\ref{eq:zmass}) must cancel 
this large contribution with a good accuracy in order 
to yield the correct $Z$-boson mass.

{}From Eq.~(\ref{eq:mhuitogsp}), we also find that this fine-tuning 
of $\mu$ 
becomes more severe if we have non-vanishing positive values of 
$m_{Q_3}^2$ and $m_{U_3}^2$ at the GUT scale.\footnote{We can think of 
introducing tachyonic squarks at the GUT scale, i.e., 
$m_{Q_3}^2$, $m_{U_3}^2 < 0$, which can reduce the fine-tuning 
as is also indicated from Eq.~(\ref{eq:mhuitogsp}). 
Such possibility has been studied in Ref.~\cite{Dermisek:2006ey}. 
In this paper, we study the fine-tuning problem without 
the tachyonic boundary conditions at the GUT scale.} 
Then, as far as the little hierarchy problem is concerned, 
it is better that the model has vanishing top squark soft 
masses at the GUT scale, 
\begin{eqnarray}
m_{Q_3}^2 &=& m_{U_3}^2 \ = \ 0, 
\label{eq:vanishingmqu}
\end{eqnarray}
and we adopt this condition in the following analysis. 

On the other hand, the Higgs soft mass squared at the GUT scale, 
$m_{H_u}^2$, appears in Eq.~(\ref{eq:mhuitogsp}) with a 
positive coefficient of $O(1)$ and then negative in 
Eq.~(\ref{eq:zmass}). Thus, $m_{H_u}^2 \sim O(M_3^2)$ 
can reduce the fine-tuning. We can approximately `renormalize' 
this contribution into the $\mu$-parameter effectively of
Eqs.~(\ref{eq:zmass}) and (\ref{eq:mutermitogsp}), 
i.e., 
\begin{eqnarray}
 & & 
-1.09 \mu^2 ~~\longrightarrow~~ -1.09 \mu^2 -0.66 m_{H_u}^2,
\label{eq:renormalize}
\end{eqnarray}
 in the 
following discussion of the fine-tuning, because the 
$m_{H_u}^2$-terms are negligible in Eqs.~(\ref{eq:mqitogsp}) 
and (\ref{eq:muitogsp}) due to the suppressed coefficients of 
$O(0.1)$. We can easily separate this effect from the effective 
$\mu$-parameter, if necessary. Then, first we just set 
\begin{eqnarray}
m_{H_u}^2 &=& 0, 
\label{eq:vanishingmh}
\end{eqnarray}
in the expressions, and consider the $\mu$-term is the 
effective one when we evaluate an effect due to 
a non-vanishing value of  $m_{H_u}^2$ at the GUT scale.

Based on these arguments, we focus on the contributions 
from $M_a$, $A_t$ and $\mu$ in Eqs.~(\ref{eq:mhuitogsp}), 
(\ref{eq:mqitogsp}) and (\ref{eq:muitogsp}) 
in the following analysis. Then, we introduce fine-tuning parameters, 
\begin{eqnarray}
\Delta_X 
&=& \frac{1}{2}\frac{X}{M_Z^2} \frac{\pa M_Z^2}{\pa X}, \qquad 
(X=\mu,\,M_1,\,M_2,\,M_3,\,A_t). 
\label{eq:ftparameter}
\end{eqnarray}
We can easily check that these parameters satisfy the relation, 
\begin{eqnarray}
\sum_X \Delta_X &=& 1, 
\label{eq:ewconstraint}
\end{eqnarray}
and then $\Delta_X \sim O(1)$ implies that the $Z$-boson mass 
is insensitive to the parameter $X$ (at the GUT scale). 
The degree of fine-tuning for the parameter $X$ can be 
considered as $100/\Delta_X$ percent.

\section{Reducing fine-tuning in bottom-up approach}
\label{sec:bottomup}
In this section, we examine the fine-tuning problem 
in a bottom-up approach, where all the soft parameters are 
regarded as independent ones to each other in their origins. 
For instance, the situation that each gauge kinetic function 
depends on different (independent) messenger fields may result 
in the independent gaugino masses at the messenger scale. 

In this case, the degree of fine-tuning in the model can 
be evaluated by the largest one $\Delta_X$ among all the fine-tuning 
parameters defined in Eq.~(\ref{eq:ftparameter}) and 
written explicitly as 
\begin{eqnarray}
\Delta_{M_1} 
&=& -0.00975 \hat{M_1}^2  
+(0.0339 \hat{M_3} 
+0.00675 \hat{M_2} 
+0.0136 \hat{A_t})\,\hat{M_1}, 
\nonumber \\
\Delta_{M_2} 
&=& -0.433 \hat{M_2}^2 
+(0.235 \hat{M_3} 
+0.00675 \hat{M_1} 
+0.0840 \hat{A_t})\,\hat{M_2}, 
\nonumber \\
\Delta_{M_3} 
&=& 5.45 \hat{M_3}^2 
+(0.0339 \hat{M_1} 
+ 0.235 \hat{M_2} 
+ 0.387 \hat{A_t})\, \hat{M_3}, 
\nonumber \\
\Delta_{A_t} 
&=& 
0.214 \hat{A_t}^2 
+(0.387 \hat{M_3} 
+0.0840 \hat{M_2} 
+0.0134 \hat{M_1})\,\hat{A_t}, 
\label{eq:deltas}
\end{eqnarray}
and 
\begin{eqnarray}
\Delta_\mu &=& -1.09 \hat\mu^2, 
\label{eq:deltamu}
\end{eqnarray}
where $\hat{M_a}=M_a/M_Z$, $\hat{A_t}=A_t/M_Z$ 
and $\hat{\mu}=\mu/M_Z$. 
When we require $|\Delta_\mu | \lesssim 10$, allowed values of 
$|\mu|$ are $|\mu | \lesssim 280$ GeV.

We easily find that $\Delta_{M_3}$ tends to be the 
largest among $\Delta_{M_a}$ and $\Delta_{A_t}$ 
for the universal gaugino masses $M_1=M_2=M_3=M$ with 
$A_t \sim O(M_a)$. In this case, if we require 
the fine-tuning for $M_3$ to be more than $10$\,\%, 
that is $\Delta_{M_3} \approx 6 \hat{M_3}^2 \le 10$, 
the gluino mass $M_3$ at the GUT scale is restricted as 
\begin{eqnarray}
M &\lesssim& 120 \ \textrm{GeV},
\qquad 
{\rm for~~}M_1=M_2=M_3=M. 
\label{eq:m3bu}
\end{eqnarray}
This does not satisfy the Higgs mass bound 
(\ref{eq:m3lb}) for $A_t(M_Z)/m_{\tilde{t}} \lesssim 1.5$. 
Therefore, a larger $A_t(M_Z) > 1.5\,m_{\tilde{t}}$ 
is inevitable as can be read off from Fig.~\ref{fig:atmt}. 
However, Eq.~(\ref{eq:atitogsp}) leads to 
$A_t(M_Z) \sim 2.8 M_3$, 
while Eq.~(\ref{eq:mttitom3}) results in 
$m_{\tilde{t}} \sim 2.2\,M_3$. 
Thus, we find $A_t(M_Z) > 1.5 m_{\tilde{t}}$ is impossible. 
On the other hand, a large value of $M_3$ like $M_3=220$ or 450 GeV 
in Eq.~(\ref{eq:m3lb}) leads to a large value of 
$\Delta_{M_3}$ like $\Delta_{M_3}=30$ or 130.
For the latter case, we need fine-tuning less than $1\%$.

\begin{figure}[t]
\begin{minipage}{0.24\linewidth}
\begin{center}
\epsfig{file=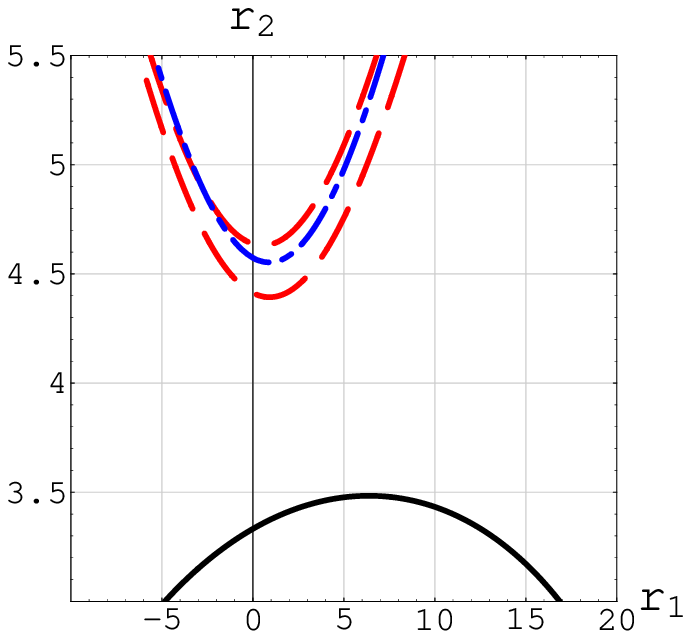,width=\linewidth} \\
$M=110$ GeV 
\end{center}
\end{minipage}
\begin{minipage}{0.24\linewidth}
\begin{center}
\epsfig{file=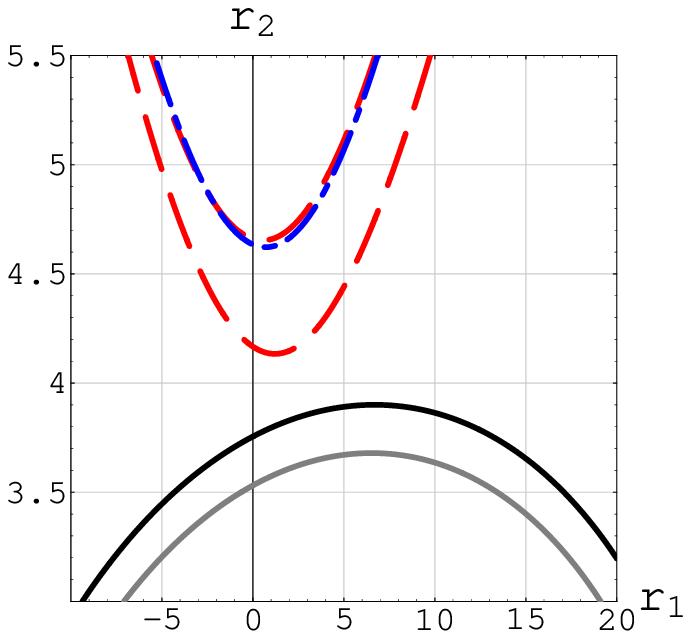,width=\linewidth} \\
$M=150$ GeV 
\end{center}
\end{minipage}
\begin{minipage}{0.24\linewidth}
\begin{center}
\epsfig{file=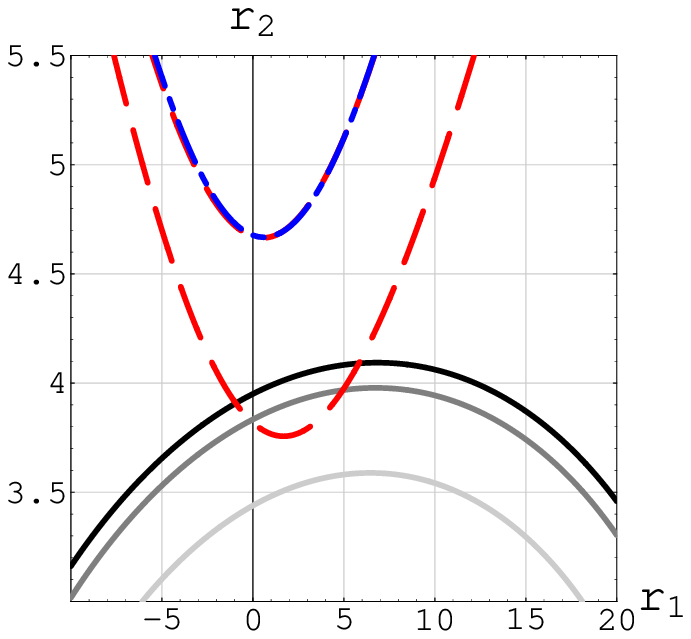,width=\linewidth} \\
$M=200$ GeV 
\end{center}
\end{minipage}
\begin{minipage}{0.24\linewidth}
\begin{center}
\epsfig{file=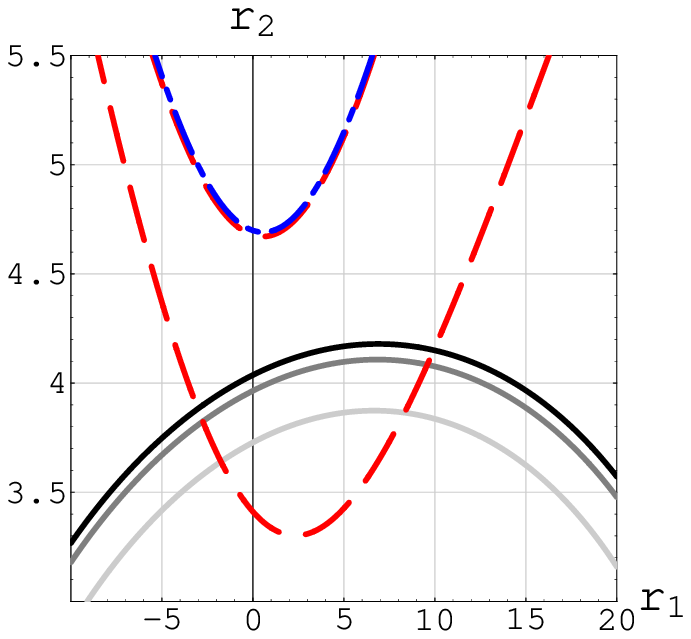,width=\linewidth} \\
$M=250$ GeV 
\end{center}
\end{minipage}
\caption{
Curves for $r_a=A_t/M_3=0$ and $m_{H_u}^2=m_{Q_3,U_3}^2=0$ 
determined by constraints from $\Delta_M=3.4,\,5,\,10$ 
(solid curves), $m_{h^0} \ge 114.4$ GeV (between two dashed curves) 
and $m_{\tilde{t}_1} \ge 95.7$ GeV (below dot-dashed curves). 
The parameter $\Delta_\mu$ is fixed by the constraint 
$\Delta_M + \Delta_\mu = 1$. 
The solid curves are darker for the smaller $\Delta_M$.}
\label{fig:r1r2raf0}
\end{figure}

The above argument for $M_1=M_2=M_3$ with $A_t \sim O(M_a)$ shows 
that only the possibility to reduce the fine-tuning associated to 
$M_3$ keeping $A_t \lesssim O(M_a)$ is {\it a departure from the 
universal gaugino mass condition} at the GUT scale.

Here we denote ratios of gaugino masses and $A_t$ by $r_1,r_2$ and 
$r_a$ as 
\begin{eqnarray}
(M_1,\,M_2,\,M_3) &=& (r_1,\,r_2,\,1)\,M, 
\qquad 
A_t \ = \ r_a\,M,
\label{eq:defr1r2m}
\end{eqnarray}
where $M$ corresponds to the overall magnitude of 
soft SUSY breaking parameters.
Note that we consider ratios, $r_1, r_2$ and $r_a$ are 
free parameters independent of $M$ in this section.
Let us define $\Delta_M$ as 
\begin{eqnarray}
\Delta_M 
&=& \sum_{a=1}^3 \Delta_{M_a} +\Delta_{A_t}. 
\label{eq:totaldm}
\end{eqnarray}
Since $\Delta_\mu = 1- \Delta_M$, we are required to obtain 
$\Delta_M \lesssim O(10)$ in order to avoid fine-tuning of $\Delta_\mu$, 
although this condition is not sufficient and small values 
$\Delta_X$ for $X=M_1,M_2,M_3$ and $A_t$ are also required.
In $\Delta_M$, the dominant contribution is due to 
$\hat {M_3}$ as obvious from Eq.~(\ref{eq:deltas}).
The next important contribution would come from 
$\hat {M_2}$, because of its sign in $\Delta_{M_2}$.
Indeed, we would obtain $\Delta_M \approx 0$ for  $r_2 \approx 4$ 
when $\hat {M_1}=A_t=0$.  
On the other hand, the $\hat {M_1}$-dependence of 
$\Delta_M$ would be small, because its coefficient is small.
This naive estimation suggests that 
the parameter region around $r_2 \sim 4$ would be favorable, 
while a larger region for $r_1$ would be favorable.

As $r_2$ increases, $m^2_{H_u}(M_Z)$ increases and 
$m^2_{U_3}(M_Z)$ decreases.
For instance, in the extremal case $r_2 \rightarrow \infty$, 
the successful electroweak symmetry breaking would not be 
realized, but the color symmetry would break radiatively. 
Thus, the parameter $r_2$ as well as others is constrained by 
experimental bounds of the stop mass and $\mu$ and the 
successful realization of electroweak symmetry breaking.

\begin{figure}[t]
\begin{minipage}{0.24\linewidth}
\begin{center}
\epsfig{file=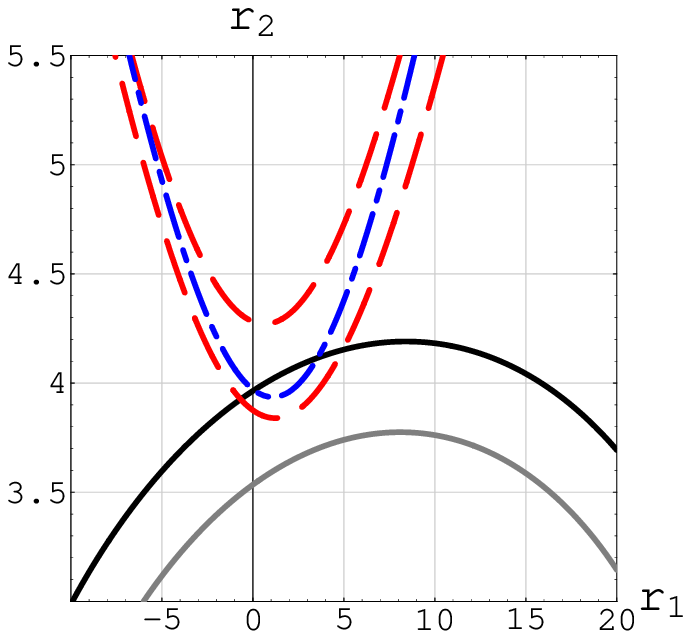,width=\linewidth} \\
$M=110$ GeV 
\end{center}
\end{minipage}
\begin{minipage}{0.24\linewidth}
\begin{center}
\epsfig{file=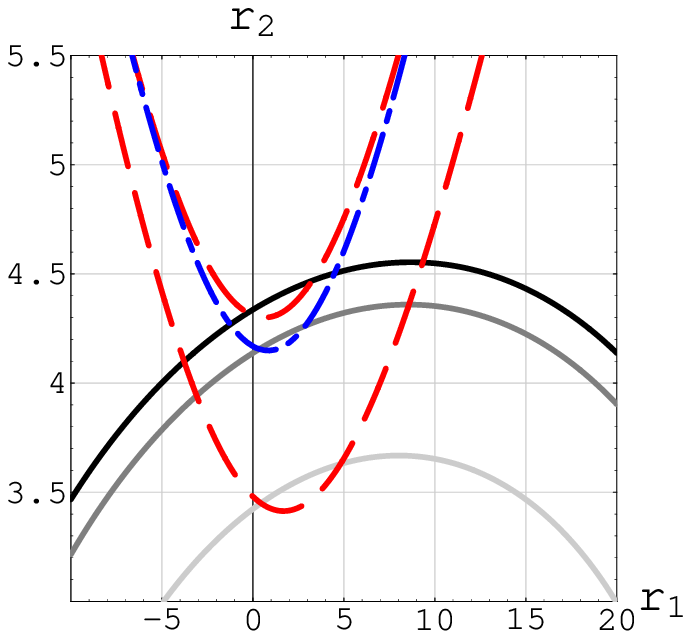,width=\linewidth} \\
$M=150$ GeV 
\end{center}
\end{minipage}
\begin{minipage}{0.24\linewidth}
\begin{center}
\epsfig{file=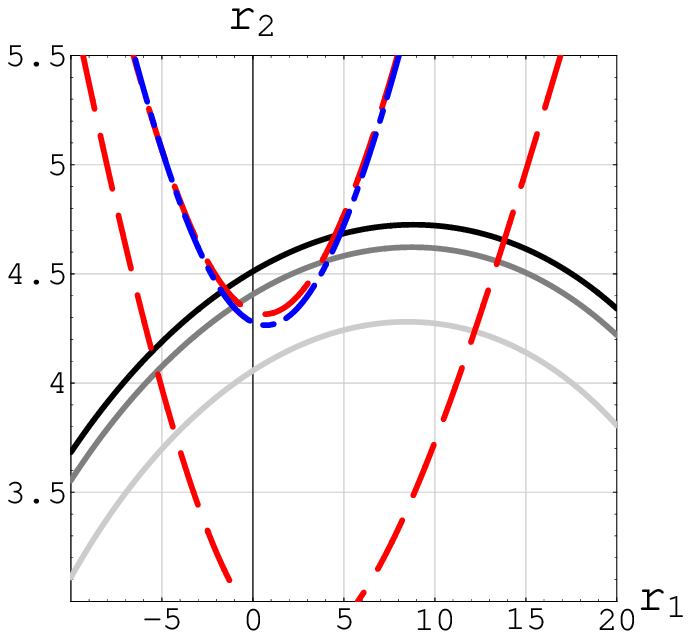,width=\linewidth} \\
$M=200$ GeV 
\end{center}
\end{minipage}
\begin{minipage}{0.24\linewidth}
\begin{center}
\epsfig{file=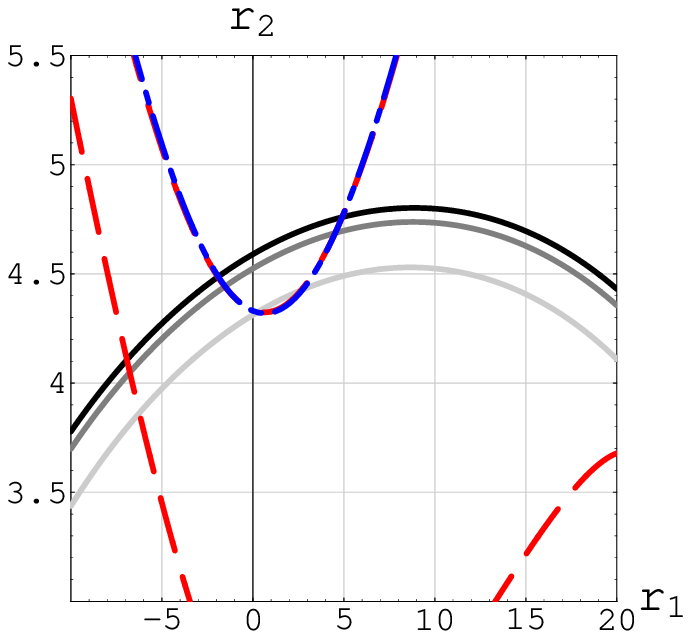,width=\linewidth} \\
$M=250$ GeV 
\end{center}
\end{minipage}
\caption{
The same curves as Fig.~\ref{fig:r1r2raf0} 
but with $r_a=A_t/M_3=1$.}
\label{fig:r1r2raf1}
\end{figure}

\begin{figure}[t]
\begin{minipage}{0.24\linewidth}
\begin{center}
\epsfig{file=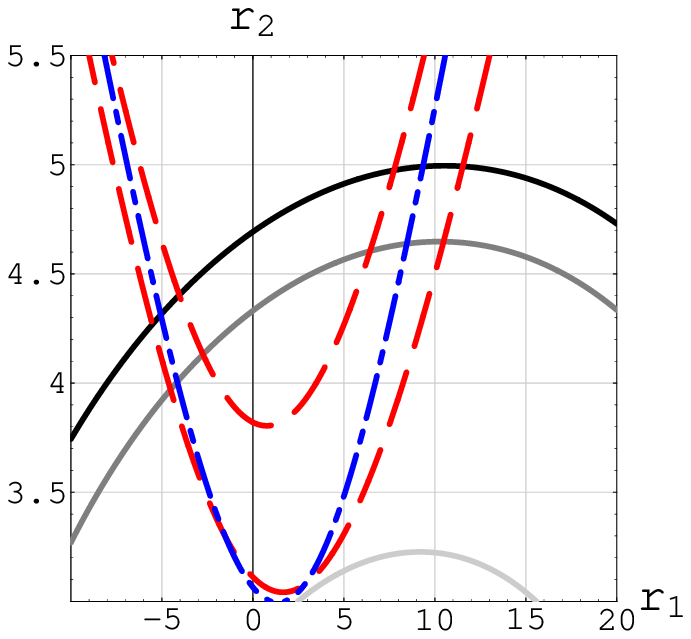,width=\linewidth} \\
$M=110$ GeV 
\end{center}
\end{minipage}
\begin{minipage}{0.24\linewidth}
\begin{center}
\epsfig{file=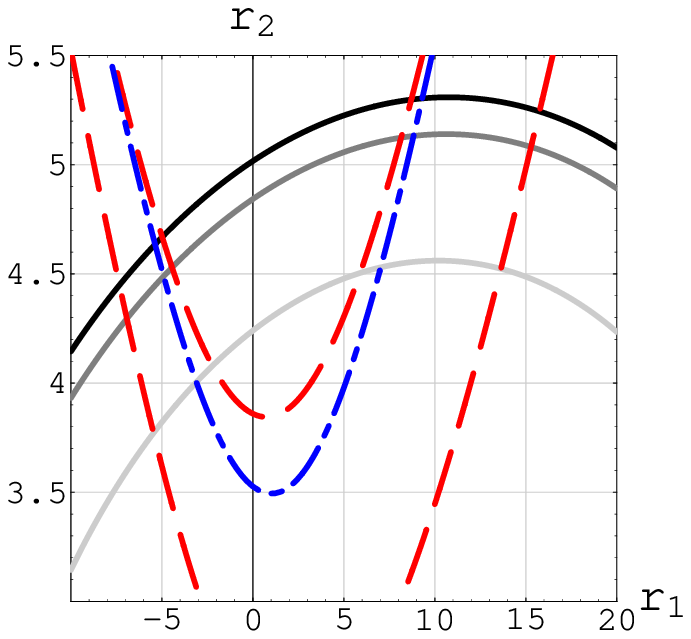,width=\linewidth} \\
$M=150$ GeV 
\end{center}
\end{minipage}
\begin{minipage}{0.24\linewidth}
\begin{center}
\epsfig{file=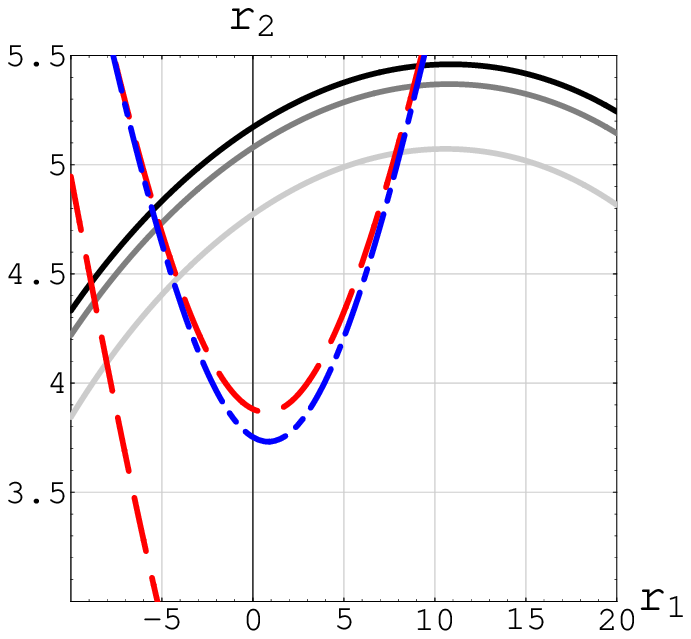,width=\linewidth} \\
$M=200$ GeV 
\end{center}
\end{minipage}
\begin{minipage}{0.24\linewidth}
\begin{center}
\epsfig{file=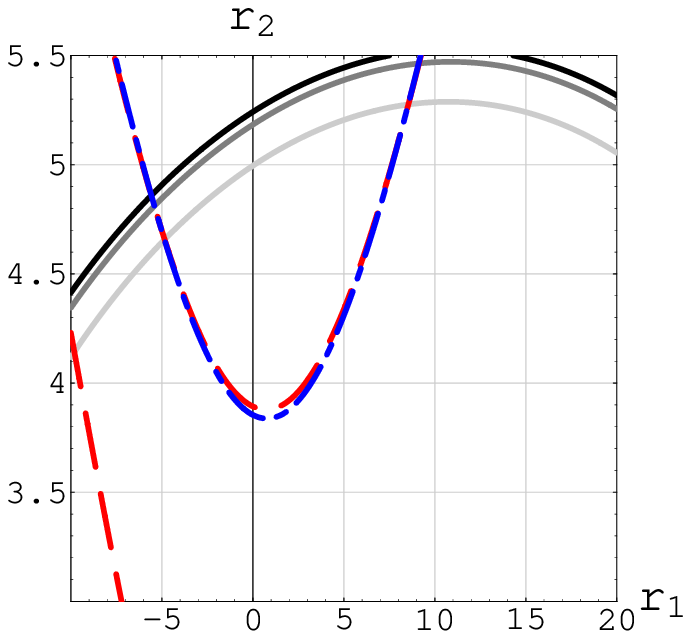,width=\linewidth} \\
$M=250$ GeV 
\end{center}
\end{minipage}
\caption{
The same curves as Fig.~\ref{fig:r1r2raf0} 
but with $r_a=A_t/M_3=2$.}
\label{fig:r1r2raf2}
\end{figure}

Figs.~\ref{fig:r1r2raf0}, \ref{fig:r1r2raf1} and 
\ref{fig:r1r2raf2} show the contours of $\Delta_M=3.4, 5, 10$ in 
$(r_1,\,r_2)$-plane for $M=110, 150, 200$ and $250$ GeV 
in the case of $r_a=0,\,1,\,2$, respectively.
The darkest and darker solid lines correspond to 
$\Delta_M=3.4$ and 5, respectively, while the less dark 
line corresponds to $\Delta_M = 10$.
Above the line corresponding to $\Delta_M=3.4$, we 
can not realize the successful electroweak symmetry breaking 
when $m_{H_u}=0$ and $|\mu(M_Z)| \geq 94$ GeV, which corresponds to 
the experimental bound of chargino mass.
%The line corresponding to $\Delta_M=1.0$ is close to 
%the line for $\Delta_M=3.4$, and above it 
%the successful electroweak symmetry breaking can not happen 
%even for $\mu(M_Z) \geq 0$ GeV and $m_{H_u}=0$, 
%although the former is not realistic from the viewpoint of 
%experimental bound of chargino mass.
%WHAT HAPPENS FOR $\Delta_M=-9$.

\begin{figure}[t]
\begin{minipage}{0.24\linewidth}
\begin{center}
\epsfig{file=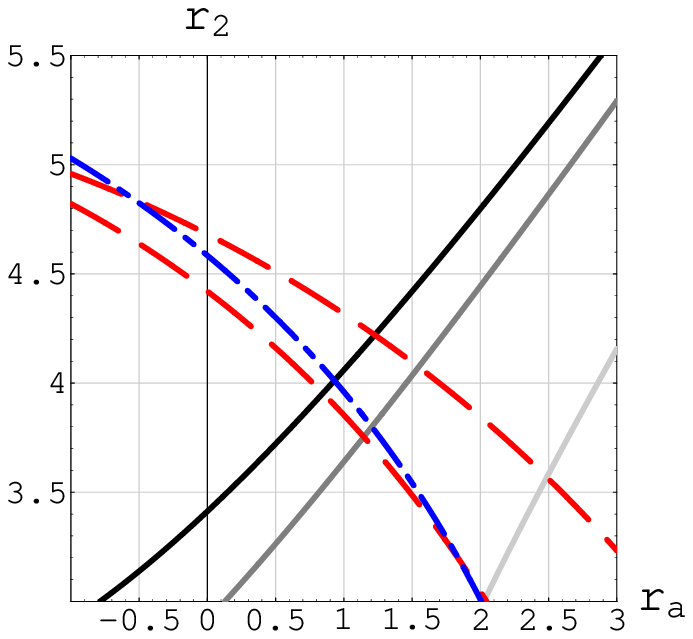,width=\linewidth} \\
$M=110$ GeV 
\end{center}
\end{minipage}
\begin{minipage}{0.24\linewidth}
\begin{center}
\epsfig{file=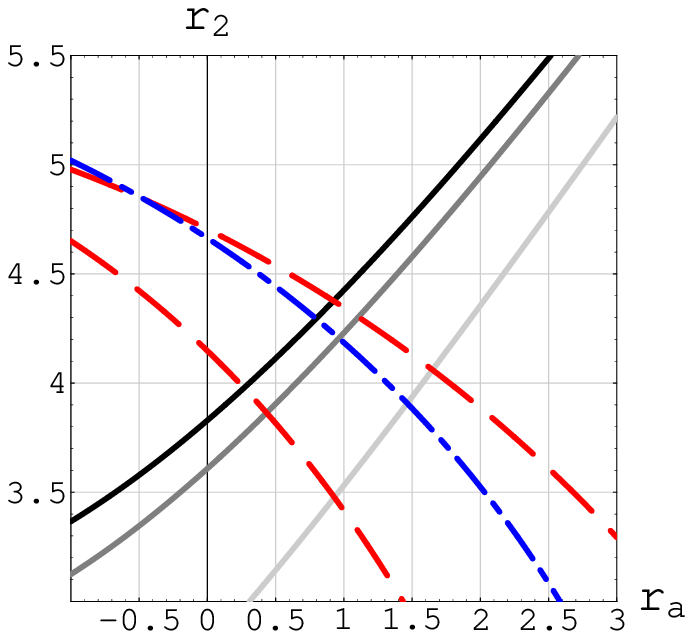,width=\linewidth} \\
$M=150$ GeV 
\end{center}
\end{minipage}
\begin{minipage}{0.24\linewidth}
\begin{center}
\epsfig{file=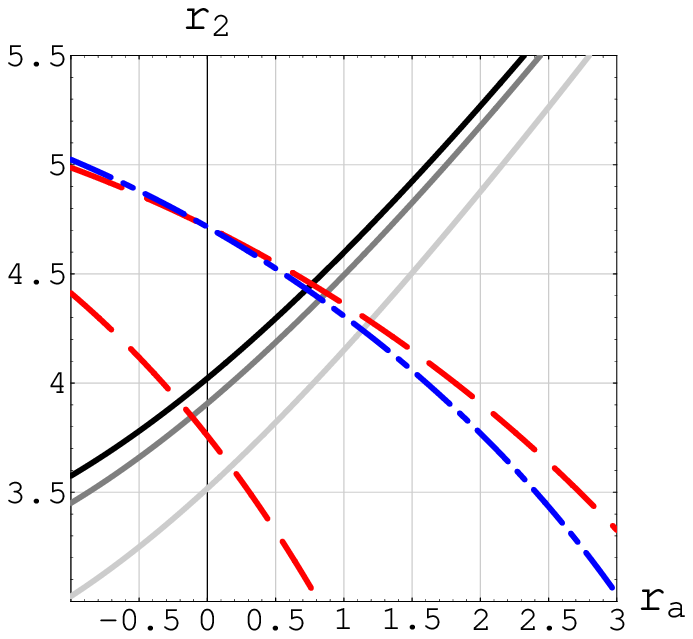,width=\linewidth} \\
$M=200$ GeV 
\end{center}
\end{minipage}
\begin{minipage}{0.24\linewidth}
\begin{center}
\epsfig{file=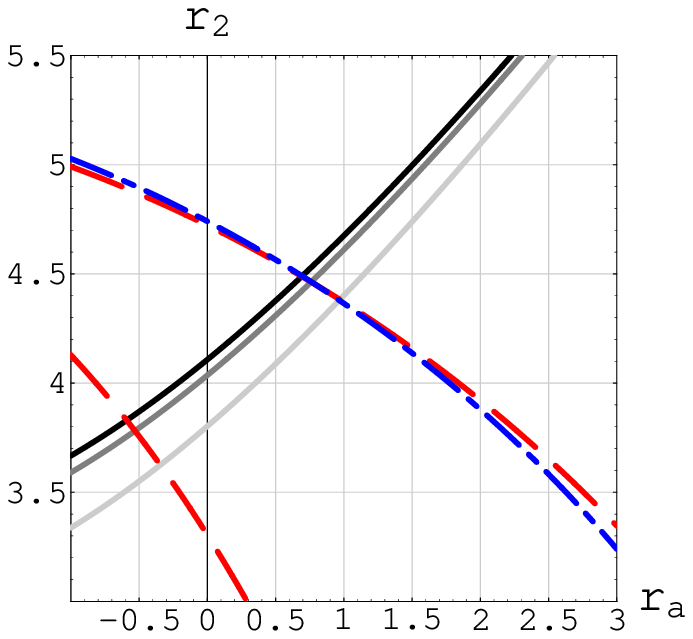,width=\linewidth} \\
$M=250$ GeV 
\end{center}
\end{minipage}
\caption{
Curves for $r_1=M_1/M_3=2$ and $m_{H_u}^2=m_{Q_3,U_3}^2=0$ 
determined by constraints from $\Delta_M=3.4,\,5,\,10$ 
(solid curves), $m_{h^0} \ge 114.4$ GeV (between two dashed curves) 
and $m_{\tilde{t}_1} \ge 95.7$ GeV (below dot-dashed curves). 
The parameter $\Delta_\mu$ is fixed by the constraint 
$\Delta_M + \Delta_\mu = 1$. 
The solid curves are more dark for the smaller $\Delta_M$.}
\label{fig:rar2r1f2}
\end{figure}

\begin{figure}[t]
\begin{minipage}{0.24\linewidth}
\begin{center}
\epsfig{file=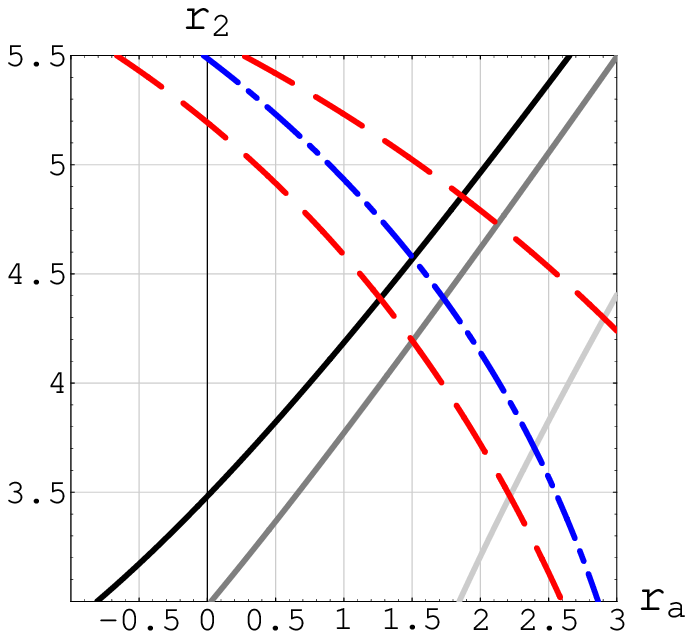,width=\linewidth} \\
$M=110$ GeV 
\end{center}
\end{minipage}
\begin{minipage}{0.24\linewidth}
\begin{center}
\epsfig{file=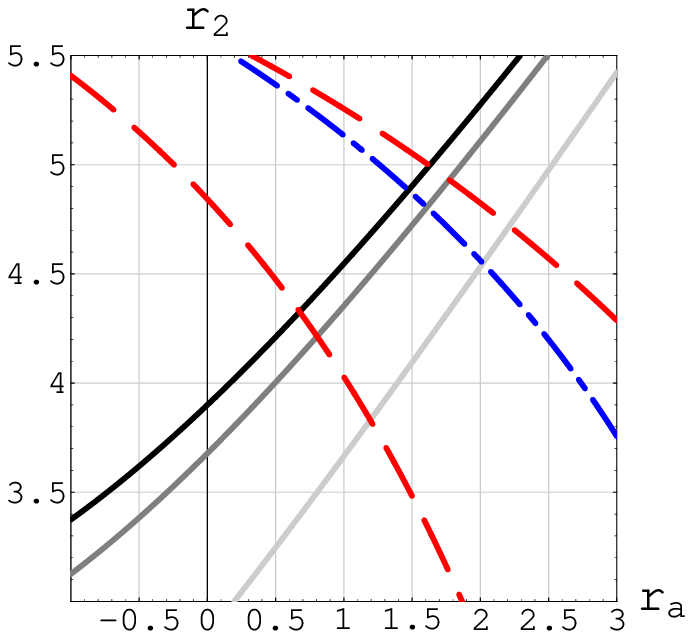,width=\linewidth} \\
$M=150$ GeV 
\end{center}
\end{minipage}
\begin{minipage}{0.24\linewidth}
\begin{center}
\epsfig{file=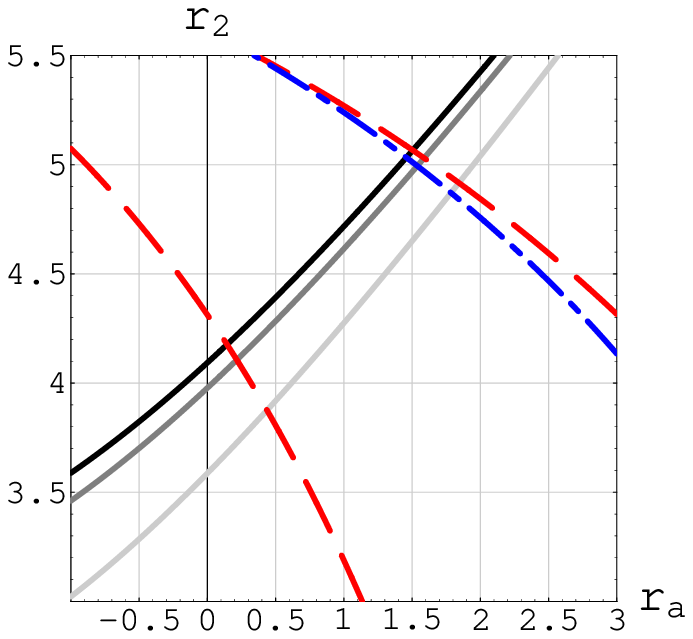,width=\linewidth} \\
$M=200$ GeV 
\end{center}
\end{minipage}
\begin{minipage}{0.24\linewidth}
\begin{center}
\epsfig{file=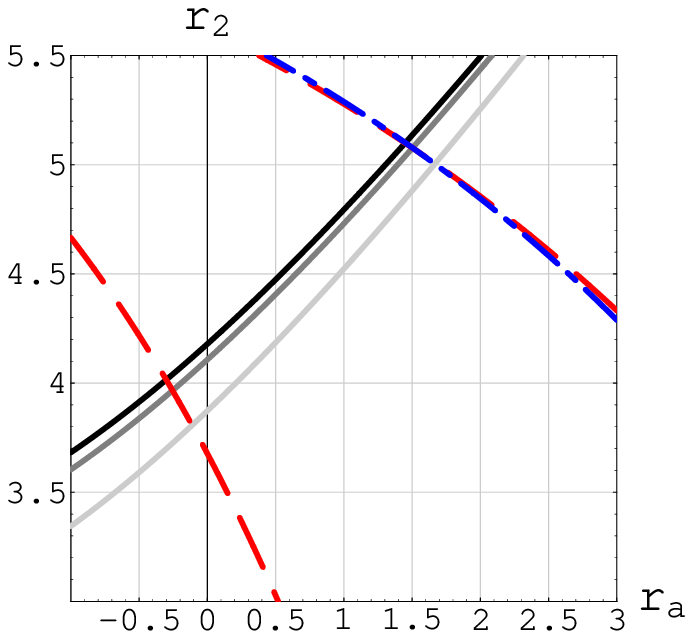,width=\linewidth} \\
$M=250$ GeV 
\end{center}
\end{minipage}
\caption{
The same curves as Fig.~\ref{fig:rar2r1f2} 
but with $r_1(M_Z)=M_1(M_Z)/M_3(M_Z)=1$ ($r_1=7.13$).}
\label{fig:rar2r1fm}
\end{figure}

In these figures, we also show the regions satisfying 
the current Higgs and top squark mass bounds~\cite{Yao:2006px}, 
$m_h \ge 114.4$ GeV and 
$m_{\tilde{t}_1} \ge 95.7$ GeV, respectively, 
where $m_{\tilde{t}_1}^2$ ($m_{\tilde{t}_2}^2$) 
is the smaller (larger) eigenvalue of 
the top squark mass-square matrix 
\begin{eqnarray}
M_{\tilde{t}}^2 &=& 
\left( 
\begin{array}{cc}
m_{Q_3}^2(M_Z)+m_t^2+\delta_Q & 
m_t \tilde{A_t} \\
m_t \tilde{A_t} & 
m_{U_3}^2(M_Z)+m_t^2+\delta_U 
\end{array}
\right), 
\nonumber
\end{eqnarray}
with 
$\delta_Q=(\frac{1}{2}-\frac{2}{3} 
\sin^2 \theta_W) \cos (2\beta) M_Z^2$, 
$\delta_U=\frac{2}{3} \sin^2 \theta_W \cos (2\beta) M_Z^2$ 
and $\sin^2 \theta_W=1-M_W^2/M_Z^2$. 
We obtain $m_h \ge 114.4$ GeV between two dashed lines, 
while we obtain $m_{\tilde{t}_1} \ge 95.7$ GeV below 
the dot-dashed line, which is close to the upper dashed line 
in several cases.
Figs.~\ref{fig:rar2r1f2}, \ref{fig:rar2r1fm} show the 
same contours as Figs.~\ref{fig:r1r2raf0}, \ref{fig:r1r2raf1} 
and \ref{fig:r1r2raf2} in $(r_a,\,r_2)$-plane for 
$M=110, 150, 200$ and $250$ GeV in the case of $r_1=2,\,7.13$, 
respectively. The ratio $r_1=7.13$ is a solution of 
$M_1(M_Z)=M_3(M_Z)$, i.e., the unification of the bino 
and the gluino mass at the EW scale.

{}From Figs.~\ref{fig:r1r2raf0}, 
\ref{fig:r1r2raf1}, 
\ref{fig:r1r2raf2}, 
\ref{fig:rar2r1f2} and 
\ref{fig:rar2r1fm}, 
we find that the Higgs mass bound 
as well as the top squark one 
is satisfied within $|\Delta_M| \le 10$ 
for the ratios $r_1$, $r_2$ and $r_a$ 
inside the region, 
\begin{eqnarray}
-10 &\lesssim& r_1 \ \lesssim \ 15, 
\qquad 
3.5 \ \lesssim \ r_2 \ \lesssim 5.5, 
\qquad 
0 \ \lesssim \ r_a \ \lesssim \ 2, 
\label{eq:allowedr1r2}
\end{eqnarray}
when the SUSY breaking scale $M$ varies from 
$110$ GeV to $200$ GeV. 
Within this region, the fine-tuning parameter $\Delta_{M_3}$ 
(contained in $\Delta_M$) given by Eq.~(\ref{eq:deltas}) 
is estimated as 
\begin{eqnarray}
5.5\,(M/M_Z)^2 
&\lesssim& 
\Delta_{M_3} 
\ \lesssim \ 
8\,(M/M_Z)^2. 
\nonumber
\end{eqnarray}
Thus, in order the fine-tuning associated to $M_3$ 
to be more than $10$\,\%, the SUSY breaking scale 
is restricted by 
\begin{eqnarray}
M &\lesssim& 110\,\textrm{-}\,120 \ \textrm{GeV}. 
\label{eq:mbu}
\end{eqnarray}
In Fig.~\ref{fig:r1r2raf0} with $r_a=0$, we find 
that there is no allowed region for $M \le 150$ GeV. 
Then, from Eq.~(\ref{eq:mbu}), we conclude that 
the non-vanishing $A$-term at the GUT scale, $r_a \ne 0$, 
is required for reducing the fine-tuning above $10$\,\% order.

\begin{table}[t]
\begin{center}
\begin{tabular}{|cc||c|c||c|c|c|}
\hline 
$M$ & (GeV) & 110 & 110 & 200 & 200 & 200 \\
\hline \hline 
$r_a$ & & 
$1$ &
$2$ &
$0$ &
$1$ &
$2$ \\
\hline 
$(r_1,\,\,r_2)$ & & 
$(3,\,\,4.0)$ &
$(10,\,\,4.8)$ &
$(2,\,\,3.9)$ &
$(5,\,\,4.6)$ &
$(10,\,\,5.4)$ \\
\hline 
$\Delta_M$ & & 
$3.8$ &
$4.3$ &
$5.1$ &
$4.7$ &
$4.4$ \\
\hline 
$\Delta_{M_3}$ & & 
$10.1$ &
$11.3$ &
$31.2$ &
$34.4$ &
$38.0$ \\
\hline 
$A_t(M_Z)/m_{\tilde{t}}$& & 
$2.0$ &
$1.9$ &
$1.6$ &
$2.4$ &
$2.2$ \\
\hline 
$M_3(M_Z)$ & (GeV) &
$321$ &
$321$ &
$583$ &
$583$ &
$583$ \\
\hline 
$M_2(M_Z)$ & (GeV) &
$361$ &
$433$ &
$640$ &
$755$ &
$886$ \\
\hline 
$M_1(M_Z)$ & (GeV) &
$135$ &
$450$ &
$164$ &
$409$ &
$818$ \\
\hline 
$\mu(M_Z)$ & (GeV) &
$108$ &
$117$ &
$130$ &
$125$ &
$120$ \\
\hline 
$m_{\tilde{t}_2}$ & (GeV) &
$436$ &
$468$ &
$714$ &
$764$ &
$820$ \\
\hline 
$m_{\tilde{t}_1}$ & (GeV) &
$202$ &
$131$ &
$247$ &
$133$ &
$186$ \\
\hline 
$m_{h,\rm{max}}$ & (GeV) &
$115$ &
$115$ &
$115$ &
$120$ &
$120$ \\
\hline
\end{tabular}
\end{center}
\caption{The mass spectra for $M=110$ and $200$ GeV 
at some typical points of $(r_1,\,r_2,\,r_a)$ 
which lead to $\Delta_M \sim 5$ ($20$\,\% tuning).}
\label{tab:spectra}
\end{table}

In Table~\ref{tab:spectra}, 
we show some mass spectra for $M=110$ GeV 
(as well as for $M=200$ GeV which will be explained later) 
at some typical points of $(r_1,\,r_2,\,r_a)$ which lead to 
$\Delta_M \sim 5$, i.e., about $20$\,\% tuning in terms of $M$. 
For $M=110$ GeV which is the marginal value of the 
condition (\ref{eq:mbu}), we find $\Delta_{M_3} \sim 10$ 
and $A_t(M_Z)/m_{\tilde{t}} \sim 2$ are realized for 
$r_a=1$ and $r_a=2$. These two are distinguished by the 
masses of the bino and the lighter top squark at the $M_Z$ scale. 
The wino mass is similar to the gluino mass for $r_a=1$, 
and is larger than it for $r_a=2$ at the $M_Z$ scale. 
This is because the larger value of $r_2$ is preferred for 
the larger value of $r_a$ in 
Fig.~\ref{fig:r1r2raf0} - Fig.~\ref{fig:rar2r1fm}. 
%WHAT HAPPENS WHEN WE VARY HIGGS AND STOP MASSES ?

Finally in this section, we summarize the discussions above. 
If all the soft parameters (as well as the $\mu$-term) at 
the GUT scale are independent to each other in their origins, 
the degree of fine-tuning in the model is almost determined 
by $\Delta_{M_3}$. A numerical evaluation indicates that 
only the possibility for relaxing the fine-tuning above $10$\,\% 
order ($\Delta_{M_3} \lesssim 10$) resides in the case of 
i) non-universal gaugino masses with the ratio inside the 
region (\ref{eq:allowedr1r2}), 
ii) a non-vanishing $A$-term at the GUT scale, $A_t>0$, 
and 
iii) a considerably low SUSY breaking scale (\ref{eq:mbu}).

\section{Fine-tuning with fixed ratios}
\label{sec:uvmodel}
It is reasonable enough to consider the situation 
that some or all of the soft SUSY breaking parameters 
share a common mass scale $M$, and the ratios between 
them are determined by some dimension less constants 
and/or geometrical numbers such as beta function 
coefficients, modular weights, and so on. 
%We can refer, e.g., Ref.~\cite{Choi:2007ka} 
%for some classifications of gaugino-mass ratios from 
%the view point of UV theories. 
Indeed, ratios of soft SUSY breaking parameters are fixed 
as certain values in each model, e.g. in 
moduli mediation, anomaly mediation, gauge messenger model and 
so on.
In this case, we do not 
need to worry about all of the fine-tuning parameters 
(\ref{eq:deltas}), and the degree of fine-tuning in 
the model is represented by only $\Delta_M$. 

In this section, we reexamine the discussions in the 
previous section, by assuming that the ratio $r_1$, 
$r_2$ and $r_a$ in Eq.~(\ref{eq:defr1r2m}) is fixed 
to some numbers by the UV theory. In this case, the 
remaining fine-tuning parameters are $\Delta_M$ and 
$\Delta_\mu$ given by Eqs.~(\ref{eq:totaldm}) and 
(\ref{eq:deltamu}), respectively. In other words, we 
worry about the sensitivity of the $Z$-boson mass to 
only the common SUSY breaking scale $M$ and the SUSY 
mass scale $\mu$. The $\Delta_{M_3}$ is a meaningless 
parameter in this sense, and thus the SUSY breaking 
scale $M$ is released from the previous upper bounds 
(\ref{eq:mbu}) or (\ref{eq:m3bu}). The numerical results 
in Fig.~\ref{fig:r1r2raf0}-Fig.~\ref{fig:rar2r1fm} show 
that $M \sim 200$ GeV possesses the widest allowed region 
of the ratios $r_1$, $r_2$ and $r_a$. This is because the 
smaller $M$ results in a smaller allowed region for the 
Higgs and the top squark mass bounds, while the region where 
$\Delta_M \le 10$ becomes narrower for the larger $M$. 
These opposite tendencies are balanced at $M \sim 200$ GeV.

\begin{figure}[t]
\begin{minipage}{0.24\linewidth}
\begin{center}
\epsfig{file=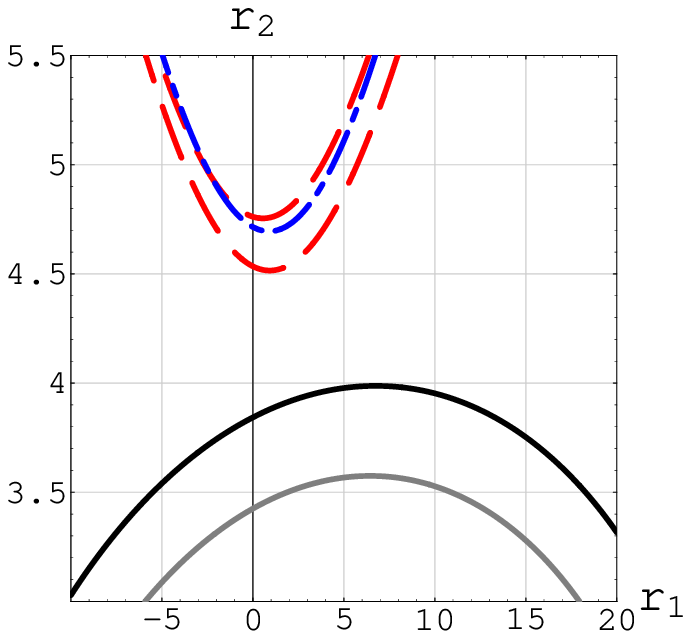,width=\linewidth} \\
$M=110$ GeV 
\end{center}
\end{minipage}
\begin{minipage}{0.24\linewidth}
\begin{center}
\epsfig{file=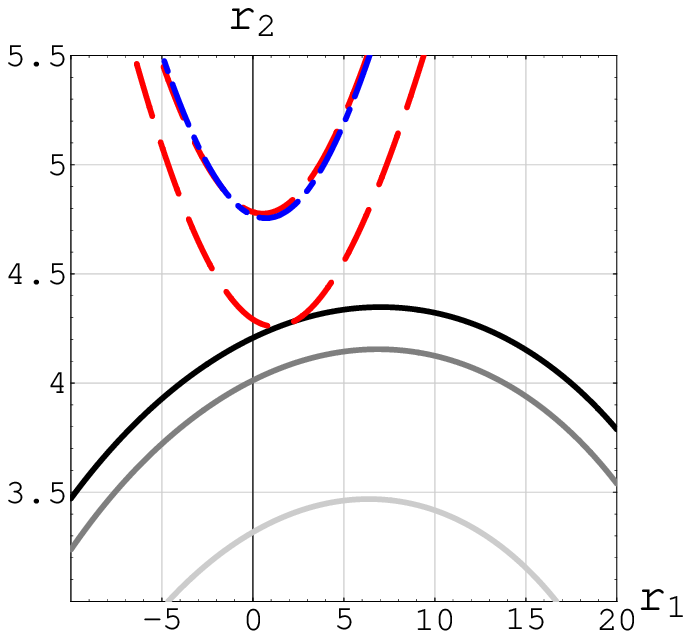,width=\linewidth} \\
$M=150$ GeV 
\end{center}
\end{minipage}
\begin{minipage}{0.24\linewidth}
\begin{center}
\epsfig{file=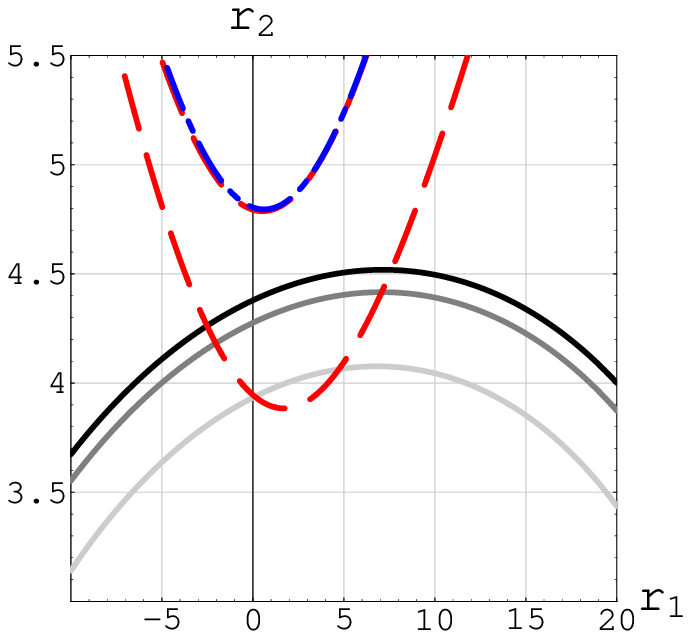,width=\linewidth} \\
$M=200$ GeV 
\end{center}
\end{minipage}
\begin{minipage}{0.24\linewidth}
\begin{center}
\epsfig{file=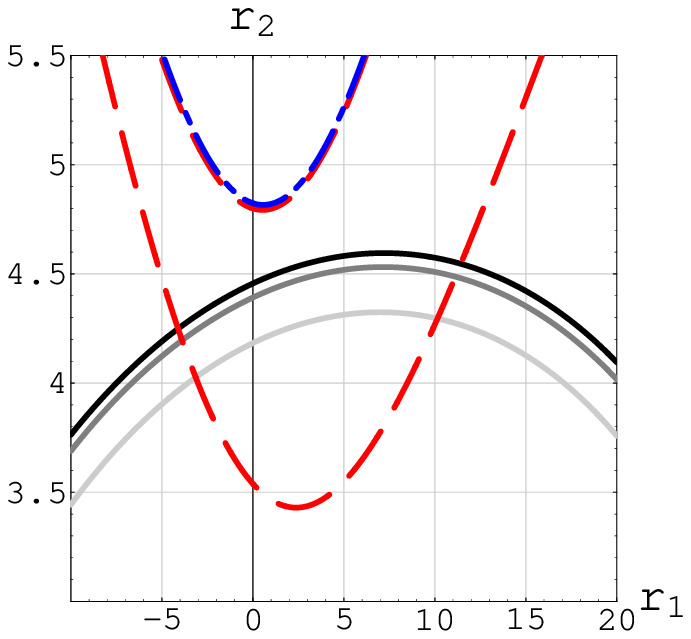,width=\linewidth} \\
$M=250$ GeV 
\end{center}
\end{minipage}
\caption{
The same curves as Fig.~\ref{fig:r1r2raf0} 
but with $m_{H_u}^2=-M^2$.}
\label{fig:r1r2raf0mh}
\end{figure}

\begin{figure}[t]
\begin{minipage}{0.24\linewidth}
\begin{center}
\epsfig{file=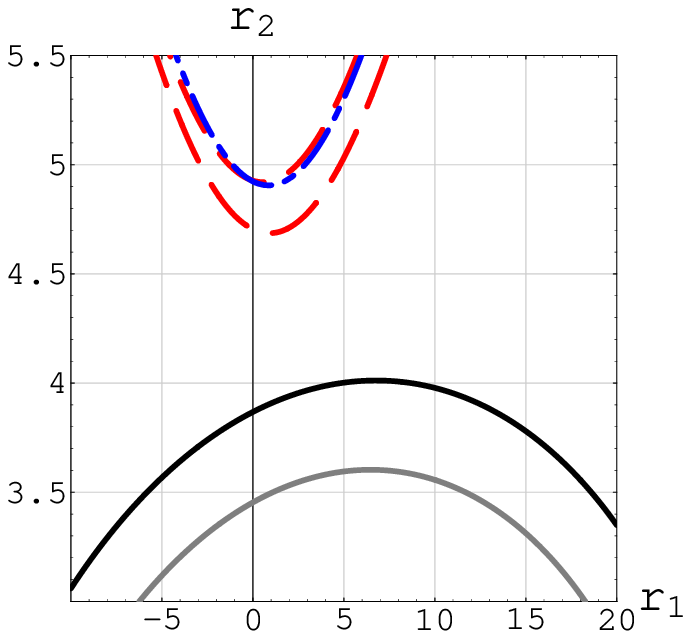,width=\linewidth} \\
$M=110$ GeV 
\end{center}
\end{minipage}
\begin{minipage}{0.24\linewidth}
\begin{center}
\epsfig{file=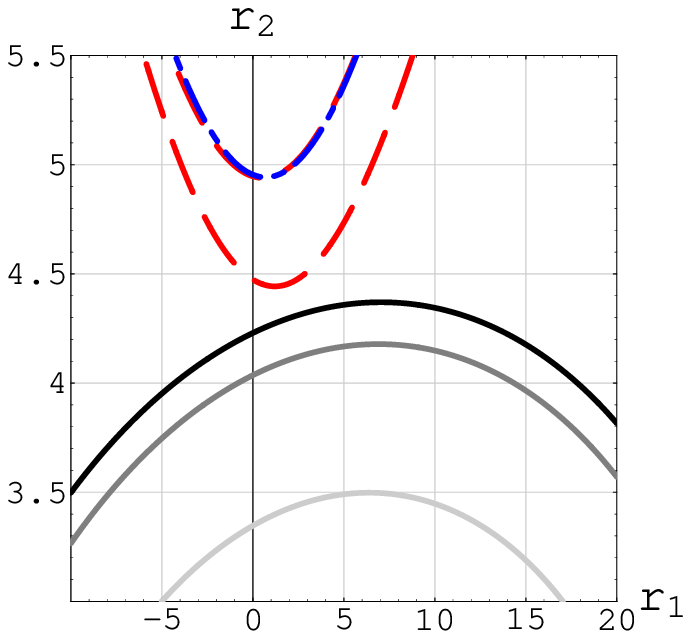,width=\linewidth} \\
$M=150$ GeV 
\end{center}
\end{minipage}
\begin{minipage}{0.24\linewidth}
\begin{center}
\epsfig{file=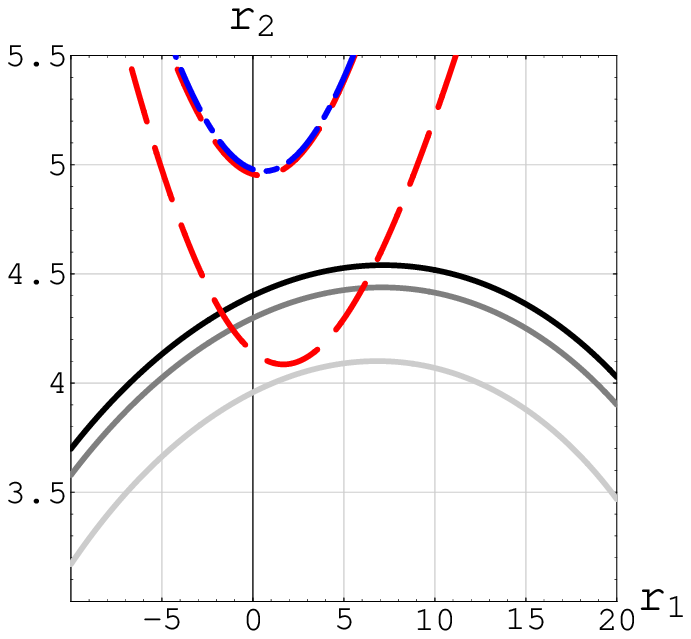,width=\linewidth} \\
$M=200$ GeV 
\end{center}
\end{minipage}
\begin{minipage}{0.24\linewidth}
\begin{center}
\epsfig{file=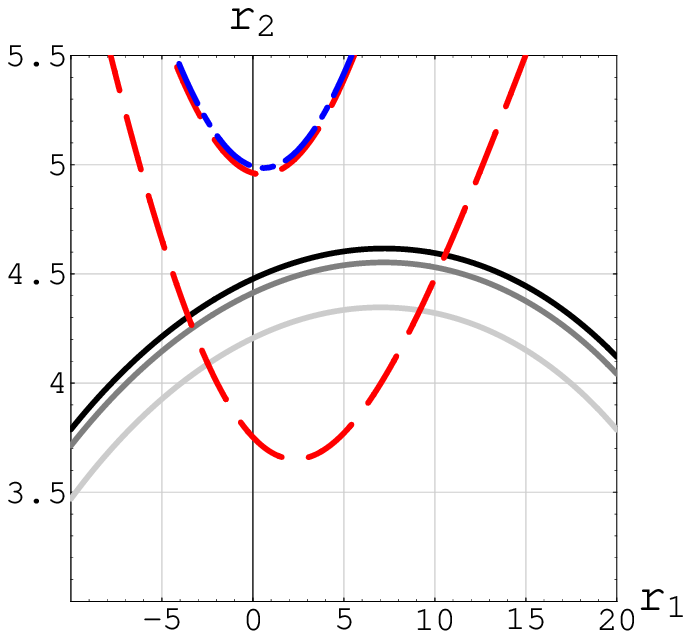,width=\linewidth} \\
$M=250$ GeV 
\end{center}
\end{minipage}
\caption{
The same curves as Fig.~\ref{fig:r1r2raf0} 
but with $m_{Q_3,U_3}^2=M^2$.}
\label{fig:r1r2raf0mqu}
\end{figure}

It is remarkable that the fine-tuning can be completely 
improved in this case. For some values of $r_1$, $r_2$ and 
$r_a$ inside (\ref{eq:allowedr1r2}), the fine-tuning parameter 
$\Delta_M$ can be of $O(1)$, and then $\Delta_\mu$ is also of 
$O(1)$ from Eq.~(\ref{eq:ewconstraint}). 
Note that $\Delta_{M_3}$ is still large $\Delta_{M_3} \gg 1$ 
for $M>120$ GeV. The point is, however, now the fine-tuning 
parameter is not $\Delta_{M_3}$ but the total sum 
$\Delta_M \sim O(1)$, where a cancellation occurs between 
$\Delta_{M_3}$ and $\Delta_{M_{1,2},A_t}$. 

At any rate, irrespective of whether we worry about 
the fine-tuning parameter $\Delta_{M_3}$ or not, 
we can obtain $\Delta_M \sim O(1)$ in the region 
(\ref{eq:allowedr1r2}) and then $\Delta_\mu \sim O(1)$. 
This implies the $Z$-boson mass is insensitive to not only 
the SUSY breaking scale $M$ but also the SUSY mass scale $\mu$. 
The small $\Delta_\mu$ corresponds to 
the small value of $\mu$ itself. It can be even the 
marginal value to the current chargino mass bound. 
%$\mu \approx 100$ GeV.
%\begin{eqnarray}
%\mu &\sim& 100 \ \textrm{GeV}. 
%\label{eq:marginalmu}
%\end{eqnarray}
%Note that $|\Delta_{\mu}| \ge 2.6$ for $\mu \ge 100$ GeV, 
%and then $\Delta_M \ge 3.4$. 
The small Higgsino mass 
is a general consequence of reduced fine-tuning 
associated to the $\mu$-parameter.

What the favored region of the ratios 
(\ref{eq:allowedr1r2}) indicates? 
First, this region is mostly close to the minimum of 
$m_{\tilde{t}}$ in terms of $A_t(M_Z)/m_{\tilde{t}}$ 
in Fig.~\ref{fig:atmt}, that is the large top squark 
mixing case~\cite{Kitano:2005wc,Dermisek:2006ey}. Second, the favored 
ratios between gaugino masses may be explained as follows. 
The region (\ref{eq:allowedr1r2}) corresponds to 
\begin{eqnarray}
-1.4 
\ \lesssim \ 
r_1(M_Z) 
\ \lesssim \ 
2.1, 
\quad 
1.0 
\ \lesssim \ 
r_2(M_Z) 
\ \lesssim \ 
1.4, 
\label{eq:allowedr1r2mz}
\end{eqnarray}
where
\begin{eqnarray}
r_1(M_Z) &=& M_1(M_Z)/M_3(M_Z) \ = \ 0.14\,r_1, 
\nonumber \\
r_2(M_Z) &=& M_2(M_Z)/M_3(M_Z) \ = \ 0.28\,r_2, 
\label{eq:r1r2mz}
\end{eqnarray}
are the gaugino mass ratios at the $Z$-boson mass scale. 
Favorable region of $r_2(M_Z)$ is rather wide, e.g. 
$\Delta r_2/r_2(M_Z)=O(0.1)$, where $r_2(M_Z) = 1.2$ and 
$\Delta r_2(M_Z) = 0.2$.
We have much wider favorable region for $r_1(M_Z)$.
That is important from the viewpoint of model building, 
because that allows $10$\% uncertainty for an explicit model.
The ratio $r_2(M_Z) \approx 1$ indicates the unification 
of the wino and the gluino masses at the EW scale. Then 
the reduced fine-tuning can be explained in the terminology 
of the so-called mirage mediation~\cite{Choi:2005uz} of 
SUSY breaking. The mirage unification of the gaugino masses 
at the EW scale~\cite{Choi:2005uz,Choi:2005hd} implies that 
the large logarithmic correction (\ref{eq:mhudominant}) 
to the $m_{H_u}^2$ is completely canceled at the EW scale 
due to the special boundary conditions at the GUT scale as 
a consequence of the mixed modulus-anomaly mediation\footnote{
In the flux compactification models~\cite{Kachru:2003aw}, the mirage 
unification scale is determined by the modulus/anomaly 
ratio of SUSY breaking mediation~\cite{Choi:2004sx}, 
which depends on the dilaton-modulus mixing ratios in the 
nonperturbative superpotential~\cite{Abe:2005rx,Choi:2006bh} 
as well as how we uplift the AdS minimum to dS 
one~\cite{Lebedev:2006qq,Choi:2006bh,Dudas:2006gr}.}. 
The range of the ratios (\ref{eq:allowedr1r2}) 
includes this type of boundary conditions as 
the central values. 

However, from (\ref{eq:allowedr1r2mz}) we find that, in order 
to reduce the fine-tuning, it is not necessary that all the 
gaugino masses are unified at the EW scale as in the mirage 
mediation models. The important one is the wino/gluino mass 
ratio, and we have a wider choice for the value of bino/gluino 
mass ratio as long as the fine-tuning is concerned. 
Inversely, the relaxed fine-tuning may predict the unification 
of the wino and gluino masses at the EW scale, but not the 
bino-gluino unification. 

In Table~\ref{tab:spectra}, 
the mass spectra for $M=200$ GeV are shown 
at some typical points of $(r_1,\,r_2,\,r_a)$ which lead to 
$\Delta_M \sim 5$, i.e., about $20$\,\% tuning in terms of $M$. 
The vanishing $A_t$ at the GUT scale $r_a=0$ is possible 
for $M=200$ GeV as well as $r_a=1,2$. In the case of $r_a=0$, 
the large $A_t(M_Z)/m_{\tilde{t}} \sim O(1)$ at the $Z$-boson 
mass scale is generated radiatively. 
The three cases $r_a=0,1,2$ are most likely distinguished by 
the mass of the bino at the $M_Z$ scale. This is due to the 
fact that the larger $r_a$ prefers the larger $r_1$ 
for $\Delta_M \le 5$ as can be seen by comparing 
Fig.~\ref{fig:r1r2raf0} - Fig.~\ref{fig:r1r2raf2}. 
The wino mass is similar to the gluino mass for $r_a=0$, 
and is larger than it for $r_a=1,2$ at the $M_Z$ scale. 
Because we are now taking such a stance that the gaugino 
masses are not independent in their origins, the value of 
$\Delta_{M_3}$ is meaningless, although it is shown in 
Table~\ref{tab:spectra} for the purpose of reference.

So far, we have considered the case with vanishing soft 
scalar masses, $m_{H_u}=m_{Q_3}=m_{U_3}=0$.
Here we comment on effects due to non-vanishing soft scalar 
masses.
First, let us evaluate effects due to non-vanishing 
value of the Higgs soft scalar mass $m_{H_u}$.
Its effect on stop masses is small.
That implies that the lightest Higgs mass $m_h$ and stop masses 
$m_{\tilde t}$ would not change significantly even when we vary $m_{H_u}$ 
in the region with $|m_{H_u}^2| \lesssim O(M^2)$.
A significant effect appears only in $m^2_{H_u}(M_Z)$, 
and such effect can be understood as `renormalization'
(\ref{eq:renormalize}).
That is, the favorable region with small $\Delta_M$ shifts 
toward the region with larger (smaller) $r_2$, when 
$m_{H_u}$ becomes negative (positive).
Fig.~\ref{fig:r1r2raf0mh} shows the case with 
$m_{H_u}^2=-M^2$.
Next, we comment on effects due to non-vanishing values of 
$m_{Q_3}$ and $m_{U_3}$.
Their effects on $\Delta_M$ are almost opposite to 
the above effect of $m_{H_u}$, because 
their signs are opposite in Eq.~(\ref{eq:mhuitogsp}).
The small $\Delta_M$ region shifts 
toward the region with larger (smaller) $r_2$, when 
$m_{Q_3}=m_{U_3}$ becomes positive (negative).
Furthermore, they also affect on the lightest Higgs mass $m_h$ and 
stop masses $m_{\tilde t}$.
Then, totally the favorable region shifts slightly when we vary 
$m_{Q_3}=m_{U_3}$, but the wideness of favorable region does not 
change drastically.
Fig.~\ref{fig:r1r2raf0mqu} shows the case with $m_{Q_3,U_3}^2=M^2$.

\section{Conclusions}
\label{sec:concl}
We studied the fine-tuning problem between the soft SUSY breaking 
parameters and the $\mu$-term for the successful electroweak symmetry 
breaking in the MSSM. The bottom-up considerations lead us to {\it the 
non-universal gaugino masses} at the GUT scale as a necessary 
condition for reducing the fine-tuning above $10$\,\% order, 
if all the soft parameters are regarded as independent ones 
to each other in their origins and no tachyonic super-particles 
are assumed at the GUT scale. In this case, the small gluino 
mass $M_3 \lesssim 120$ GeV and the non-vanishing $A$-terms 
$A_t>0$ at the GUT scale is required from $\Delta_{M_3} \lesssim 10$. 

On the other hand, if the soft SUSY breaking parameters share 
a common mass scale $M$ with the fixed ratios by the UV theory, 
each fine-tuning parameter such as $\Delta_{M_3}$ does not make 
any sense. Only the total one such as 
$\Delta_M=\sum_{a=1}^3 \Delta_{M_a}+\Delta_{A_t}$ as well as the 
SUSY parameter $\Delta_\mu$ represents the degree of 
fine-tuning in the model. In this case, the above upper-bound 
on $M_3$ disappears, and then we find the fine-tuning can be 
completely improved in some models of non-universal gaugino masses. 
A numerical evaluation shows that the model with the gluino mass 
$M_3 \sim 200$ GeV at the GUT scale has the widest allowed 
range of $r_1=M_1/M_3$, $r_2=M_2/M_3$ and $r_a=A_t/M_3$. 
In this case of the least fine-tuning, even the vanishing $A_t$ 
at the GUT scale is possible and a relatively large 
$A_t(M_Z)/m_{\tilde{t}}>1.5$ at the $Z$-boson mass scale 
is generated radiatively. 

In both the above approaches, the non-universal gaugino mass 
conditions, especially, $M_2 \approx 4 M_3$ at the GUT scale 
is the key to improve the fine-tuning. 
This implies the wino and gluino degeneracy at the weak scale. 
Another implication is a smaller Higgsino mass due to the 
reduced or eliminated fine-tuning $\Delta_M \le 10$ accompanying 
$|\Delta_\mu| \le 10$. The bino mass $M_1$ at the GUT scale 
is less constrained from our discussions of fine-tuning. 
This fact implies that the EW mirage-unification 
model~\cite{Choi:2005uz,Choi:2005hd}, 
where all the gaugino masses are unified at the EW scale, 
can be deformed such that only the wino and gluino masses are 
unified, keeping the absence of fine-tuning. In other words, 
the $U(1)_Y$ gauge kinetic function can have a different 
origin from the other ones for $SU(3)_C$ and $SU(2)_L$. 
It would be important to study model building at high energy scale, 
extending the low-energy scale mirage \cite{Choi:2005uz,Choi:2005hd}.
We would study elsewhere explicit construction of 
such partial mirage model, where only the gluino and wino masses are 
degenerate around $M_Z$.
A negative value of $m_{H_u}$ makes the favorable region wider, and 
larger value of $M_2/M_3$ becomes favorable.
On the other hand, when we vary stop masses $m_{Q_3,U_3}$, 
the situation does not change drastically.

Our favorable value of $\mu$ is small.
For example we have $|\mu| \lesssim 280$ GeV, when we require 
$\Delta_\mu \lesssim 10$.
In addition, the bino mass $M_1$ can vary in a quite wide range.
This aspect would be interesting from the viewpoint of 
dark matter candidate.

We have concentrated to the Higgs sector and the electroweak 
symmetry breaking, to which only gaugino masses, stop masses and 
Higgs masses are relevant.
Other mass parameters are irrelevant to our discussion, 
that is, they can be more model-dependent.

\subsection*{Acknowledgement}
The authors would like to thank Haruhiko Terao for useful discussions.
H.~A.\/ and T.~K.\/ are supported in part by the Grand-in-Aid for 
Scientific Research \#182496, \#17540251, respectively. 
T.~K.\/ is also supported in part by the Grant-in-Aid for the 21st Century 
COE ``The Center for Diversity and Universality in Physics'' from the 
Ministry of Education, Culture, Sports, Science and Technology of Japan.

\end{document}